\documentclass[11pt]{article}

\usepackage{latexsym}
\usepackage{amsmath}
\usepackage{amssymb}
\usepackage{xspace}

\usepackage{epic,eepic}
\usepackage{graphicx}
\usepackage{color}
\definecolor{Gray}          {cmyk}{0,0,0,0.50}
\definecolor{Black}         {cmyk}{0,0,0,1}
\definecolor{White}         {cmyk}{0,0,0,0}
\definecolor{LightGray}     {cmyk}{0,0,0,0.20}
\newtheorem{xdefinition}{Definition}
\newtheorem{xobservation}{Observation}
\newtheorem{xtheorem}{Theorem}
\newtheorem{xlemma}{Lemma}
\newtheorem{xproposition}{Proposition}
\newtheorem{xcorollary}{Corollary}
\newenvironment{definition}{\begin{xdefinition}\rm}%
{\hspace*{\fill}\raisebox{-1pt}{\boldmath$\Box$}\end{xdefinition}}
{\hspace*{\fill}\raisebox{-1pt}{\boldmath$\Box$}\end{xobservation}}
\newenvironment{theorem}{\begin{xtheorem}\rm}{\end{xtheorem}}
\newenvironment{lemma}{\begin{xlemma}\rm}{\end{xlemma}}
\newenvironment{proposition}{\begin{xproposition}\rm}{\end{xproposition}}
\newenvironment{corollary}{\begin{xcorollary}\rm}{\end{xcorollary}}
\newenvironment{proof}{\begin{trivlist}\item[]{\bf Proof }}%
{\hspace*{\fill}\raisebox{-1pt}{\boldmath$\Box$}\end{trivlist}}

\newcommand{\MIN}[1]{\min\left\{#1\right\}}

\newcommand{\FLOOR}[1]{\left\lfloor#1\right\rfloor}
\newcommand{\SIZE}[1]{|#1|}
\newcommand{\ALG}{\alg{A}}
\newcommand{\ALGB}{\alg{B}}
\newcommand{\LAZY}{{\ensuremath{\textsc{Lazy}}}\xspace}
\newcommand{\LALG}{{\ensuremath{\cal L\ALG}}\xspace}
\newcommand{\OPT}{{\ensuremath{\textsc{Opt}}}\xspace}
\newcommand{\DC}{{\ensuremath{\textsc{Dc}}}\xspace}
\newcommand{\aDC}{{\ensuremath{a\textsc{-Dc}}}\xspace}
\newcommand{\LDC}{{\ensuremath{\textsc{Ldc}}}\xspace}
\newcommand{\aLDC}{{\ensuremath{a\textsc{-Ldc}}}\xspace}
\newcommand{\bLDC}{{\ensuremath{b\textsc{-Ldc}}}\xspace}
\newcommand{\iaLDC}{{\ensuremath{\frac{1}{a}\textsc{-Ldc}}}\xspace}
\newcommand{\Greedy}{{\ensuremath{\textsc{Greedy}}}\xspace}
\newcommand{\BAL}{{\ensuremath{\textsc{Bal}}}\xspace}
\newcommand{\MOD}[2]{#1~\text{mod}~#2}
\newcommand{\alg}[1]{{\ensuremath{\mathbb{#1}}}\xspace}
\newcommand{\worst}[1]{\ensuremath{#1_{\text{W}}(I)}\xspace}
\newcommand{\clower}{\ensuremath{c_{\text{l}}(\alg{A},\alg{B})}\xspace}
\newcommand{\cupper}{\ensuremath{c_{\text{u}}(\alg{A},\alg{B})}\xspace}
\newcommand{\rwor}[2]{\ensuremath{\text{WR}_{#1,#2}}\xspace}
\newcommand{\cupperba}{\ensuremath{c_{\text{u}}(\alg{B},\alg{A})}\xspace}
\newcommand{\DAF}{\frac{d}{a}}
\newcommand{\DBF}{\frac{d}{b}}
\newcommand{\DOWN}[2]{[#1]_{#2}}
\newcommand{\SET}[1]{\{#1\}}
\newcommand{\bigbinom}[2]{\left(\!\begin{array}{c}#1\\#2\end{array}\!\right)}
\newcommand{\WEHAVE}{\!:\;}

\begin{document}

\title{A Comparison of \\ Performance Measures for Online Algorithms\thanks{A preliminary version 
 of this paper appeared in {\em  11th International Algorithms and
Data Structures Symposium (WADS 2009)}, volume 5664 of {\em Lecture
Notes in Computer Science}, pages
119-130, Springer, 2009.}
}

\author{Joan Boyar\thanks{Department of Mathematics and Computer Science,
University of Southern Denmark, Campusvej 55, DK-5230 Odense~M, Denmark,
\{joan,kslarsen\}@imada.sdu.dk.
Supported in part by
the Danish Council for Independent Research.
Part of this work was carried out while these authors
were visiting the University of California, Irvine, and the University of
Waterloo, Canada.
}
\and
Sandy Irani\thanks{Department of Computer Science, University of California, Irvine,
CA 92697, USA,
irani@ics.uci.edu.
Supported in part by NSF Grants CCR-0514082 and CCF-0916181.}
\and
\addtocounter{footnote}{-2}
Kim S. Larsen\footnotemark
}

\maketitle

\begin{abstract}
This paper provides a systematic study of several proposed
measures for online algorithms in the context of a specific problem,
namely, the two server problem on three colinear points.
Even though the problem is simple, it encapsulates a core challenge in
online algorithms which is to balance greediness and adaptability.
We examine Competitive Analysis, the Max/Max Ratio, the Random Order Ratio,
Bijective Analysis and Relative Worst Order Analysis, and determine
how these measures compare the Greedy Algorithm, Double Coverage,
and Lazy Double Coverage,
commonly studied algorithms in the context of server problems.
We find that by the Max/Max Ratio and Bijective Analysis, Greedy is the
best of the three algorithms. Under the other measures, Double Coverage and
Lazy Double Coverage are
better, though Relative Worst Order Analysis indicates
that Greedy is sometimes better.
Only Bijective Analysis and Relative Worst Order Analysis indicate that
Lazy Double Coverage is better than Double Coverage.
Our results also provide the first 
proof of optimality of an algorithm under Relative Worst Order Analysis.
\end{abstract}

\section{Introduction}

Since its introduction by Sleator and Tarjan in 1985~\cite{ST85},
Competitive Analysis has been the most widely used method for evaluating 
online algorithms. A problem is said to be {\em online} if the input to
the problem is given a piece at a time, and the algorithm must commit to parts
of the solution over time before the entire input is revealed to the algorithm.
{\em Competitive Analysis} evaluates an online algorithm in comparison to 
the optimal offline algorithm which receives the input in its entirety in advance
and has unlimited computational power in determining a solution.
Informally speaking, one considers the worst-case input which maximizes the
ratio of the cost of the online algorithm for that input to the cost
of the optimal offline algorithm
on that same input. The maximum ratio achieved is
called the {\em Competitive Ratio}.
Thus, one factors out the inherent difficulty
of a particular input (for which the offline algorithm
is penalized along with the online algorithm)
and measures what is lost in making decisions with partial information
and/or limited power.

Despite the popularity of Competitive Analysis, researchers have been well
aware of its deficiencies and have been seeking better alternatives
almost since the time that it came into wide use. (See~\cite{DLO05} for
a fairly recent survey.)
Many of the problems with Competitive Analysis stem from the fact that
it is a worst case measure and fails to examine the performance of
algorithms on instances that would be expected in a particular application.
It has also been observed
that Competitive Analysis sometimes fails to distinguish between
algorithms which have very different performance in practice
and intuitively differ in quality.

Over the years, researchers have devised alternatives to Competitive
Analysis, each designed to address one or all of its shortcomings.
There are exceptions, but it is fair to say that many alternatives
are application-specific, and very often, the papers 
in which they are introduced only present a direct
comparison between a new measure and Competitive Analysis.

This paper is a study of several generally-applicable
alternative measures for evaluating
online algorithms that have been suggested in
the literature. We perform this
comparison in the context of a particular problem: the 2-server
problem on the line with three possible request points,
nick-named here the {\em baby server problem}.
Investigating simple $k$-server problems to shed light on new ideas
has also been done in~\cite{BIK08}, for instance.

We focus on three algorithms, \Greedy{}, {\sc Double Coverage}
(\DC)~\cite{CKPV91}, and
{\sc Lazy Double Coverage} (\LDC),
and four different analysis techniques (performance measures):
Bijective Analysis, the Max/Max Ratio, the Random Order Ratio, and Relative
Worst Order Analysis.

In investigating the baby server problem, we find that according to
some quality measures for online algorithms,
\Greedy{} is better than \DC{} and \LDC{}, whereas for others,
\DC and \LDC{} are better than \Greedy{}. In addition, for some measures
\LDC{} is better than \DC{}, while for others they are indistinguishable.

The analysis methods that conclude that \DC{} and \LDC{} are better than
\Greedy{} are focused on a 
worst-case
sequence for the ratio of an algorithm's cost compared to \OPT{}.
In the case of \Greedy{} vs.\ \DC{} and \LDC{}, this conclusion 
makes use of the fact that
there exists a family of sequences for which \Greedy{}'s cost is unboundedly
larger than the cost of \OPT{}, whereas for each of \DC{} and \LDC{}, the cost
is always at most a factor of two larger than the cost of \OPT{}.

On the other hand, the measures that conclude that \Greedy{} is best compare
two algorithms based on the multiset of costs stemming from 
the set of all sequences
of a fixed length.
In the case of \Greedy{} and \LDC{}, this makes use
of the fact that for any fixed $n$,
both the maximum as well as the average cost of \LDC{} over all
sequences of length~$n$ are greater than the corresponding values
for \Greedy{}.

Using Relative Worst Order Analysis a more nuanced result is obtained,
concluding that \LDC{} can be  at most a factor of two worse than \Greedy{},
while \Greedy{} can be unboundedly worse than \LDC{}.

The analysis methods that distinguish between \DC{} and \LDC{}
(Bijective Analysis and Relative Worst Order Analysis) take
advantage of the fact that \LDC{} performs at least as well as \DC{}
on every sequence and performs better on some. The others (Competitive
Analysis, the Max/Max Ratio, and the Random Order Ratio) cannot
distinguish between them, due to the intermediate
comparison to \OPT{}, i.e., algorithms are compared to \OPT{} and
then the results of this comparison are compared. On some sequences
where \DC{} and \LDC{} do worst in comparison with \OPT{}, they
perform identically, so these worst case measures conclude that the
two algorithms perform identically overall. This phenomenon occurs
in other problems also. For example, some analysis methods
fail to distinguish between the paging algorithms LRU and FWF, even
though the former is clearly better and is at least as good on every
sequence.

The simplicity of the baby server problem also
enables us to give the first proof of
optimality in Relative Worst Order Analysis:
\LDC{} is an optimal algorithm for this problem.

Though our main focus is the greediness/adaptability issue that
we investigate through the analyses of \Greedy{} and \LDC{} over
a broad collection of quality measures, we also include some
results about the balance algorithm~\cite{MMS90}, \BAL{}.
Because of the interest for this server algorithm in the literature,
we find it natural to mention the results for \BAL{} that can
be obtained relatively easily within our framework.

\section{Preliminaries}
\label{preliminaries}
In this section,
we define the server problem used throughout this
paper as the basis for our comparison.
We also define the server algorithms used, and the quality measures which
are the subject of this study.

\subsection{The Server Problem}
Server problems~\cite{BE97} have been the objects of many studies.
In its full generality, one assumes that some number $k$
of servers are available in some metric space.
Then a sequence of requests must be treated.
A request is simply a point in the metric space,
and a $k$-server algorithm must move servers in response
to the request to ensure that at least one server is
placed on the request point. A cost is associated with
any move of a server (this is usually the distance moved
in the given metric space), and the objective is to minimize
total cost.
The initial configuration (location of servers) may or
may not be a part of the problem formulation.

In investigating the strengths and weaknesses of the various
measures for the quality of online algorithms, we define
the simplest possible nontrivial server problem:

\begin{definition}
The {\em baby server problem} is a 2-server problem on the line
with three possible request points $A$, $B$, and $C$,
in that order from left to right,
with distance one between $A$ and $B$ and integral distance $d \geq 2$
between $B$ and $C$.
The cost of moving a server is defined to be the distance it is moved.
We assume that initially the two servers are placed on $A$ and $C$.
\end{definition}

As a side remark, we have considered most proofs in this paper
in the context of a non-integral distance $d$ between $B$ and $C$.
The main conclusions remain the same, but many of the proofs become
longer and the formulas less readable.
In a few places, we consider variants of \LDC, where the right-most server
moves at a speed $a$ times faster than the left-most server.
Also in this case we assume that $d/a$ is integral in order to highlight
the core findings.

All results in the paper pertain to the baby server problem.
Even though the problem is simple, it requires
balancing greediness and adaptability which is a central
problem in all $k$-server settings and many online problems in general.
This simple problem we consider is
sufficient to show the non-competitiveness of \Greedy{} with respect to
Competitive Analysis~\cite{BE97}.

\subsection{Server Algorithms}
First, we define some relevant properties of server algorithms:
\begin{definition}
A server algorithm is called
\begin{itemize}
\item {\em noncrossing} if servers never change their
relative position on the line.
\item {\em lazy}~\cite{MMS90} if it never moves
more than one server in response to a request and it does not
move any servers if the requested point is already occupied by
a server.
\end{itemize}
A server algorithm fulfilling both these properties is
called {\em compliant}.
\end{definition}

Given an algorithm, \ALG, we define the algorithm {\em lazy} \ALG,
\LALG, as follows: \LALG{} will maintain a {\em virtual} set of servers and 
their locations as well as the real set of
servers in the metric
space. There is a one-to-one correspondence between
real servers and virtual servers.
The virtual set will simulate the behavior of \ALG.
The initial server positions of the virtual and real servers
are the same.

When a request arrives, the  virtual servers are moved in accordance 
with algorithm \ALG.  After this happens, there will always be at least one 
virtual server on the requested point. Then the real servers move to 
satisfy the request: If there is already a real server on the requested 
point, nothing more happens. Otherwise, the real server corresponding to 
the virtual server on the requested point moves to the requested point. 
If there is more than one virtual server on the requested point, 
tie-braking rules may be applied. In our case, we will pick the closest 
server to move to the requested point.

General $k$-server problems that are more complicated than the
baby server problem may need more involved tiebreaking rules to be
deterministically defined.
Note that as a special case of the above, a virtual move can be
of distance zero, while still leading to a real move of non-zero distance.

In~\cite{CKPV91}, it was observed that
for any 2-server algorithm, there exists a noncrossing
algorithm with the same cost on all sequences.
In~\cite{MMS90}, it was observed that
for an algorithm \ALG{} and its lazy version \LALG{},
for any sequence $I$ of requests, $\ALG{}(I) \geq \LALG{}(I)$
(we refer to this as the {\em laziness observation}).
Note that the laziness
observation applies to the general $k$-server problem, so the
results that depend only on this observation can also be generalized beyond the
baby server problem.

We define a number of algorithms by specifying  their behavior
on the next request point, $p$.
For all algorithms considered here, no moves are made if a server already occupies
the request point (though internal state changes are sometimes made
in such a situation).

\Greedy{} moves the closest server to $p$. Note that due to the problem
formulation, ties cannot occur (and the server on $C$ is never moved).

If $p$ is in between the two servers, Double Coverage (\DC), moves both
servers at the same speed in the direction of $p$
until at least one server reaches the point. If $p$ is on the same
side of both servers, the nearest server moves to $p$.

We define \aDC{} to work in the same way as \DC{}, except that
the right-most server moves at a speed $a\leq d$ times faster than
the left-most server.
We refer to the lazy version of \DC{} as \LDC{}
and the lazy version of \aDC{} as \aLDC{}.

The balance algorithm~\cite{MMS90}, \BAL, makes its decisions based
on the total distance travelled by each server.
For each server, $s$, let $d_s$ denote the total distance travelled
by $s$ from the initiation of the algorithm up to the current point
in time.
On a request, \BAL{} moves a server, aiming to obtain the smallest
possible $\max_s d_s$ value {\em after} the move.
In case of a tie, \BAL{} moves the server which must move
the furthest.

\begin{figure}[htp]
\begin{center}
\begin{picture}(335,166)
\put(75,0){\begin{picture}(260,166)(0,0)
\put(20,40){\line(0,-1){5}}
\put(20,20){\makebox(0,0){$0$}}
\put(40,40){\line(0,-1){5}}
\put(40,20){\makebox(0,0){$1$}}
\put(60,40){\line(0,-1){5}}
\put(60,20){\makebox(0,0){$2$}}
\put(80,40){\line(0,-1){5}}
\put(80,20){\makebox(0,0){$3$}}
\put(100,40){\line(0,-1){5}}
\put(100,20){\makebox(0,0){$4$}}
\put(120,40){\line(0,-1){5}}
\put(120,20){\makebox(0,0){$5$}}
\put(140,40){\line(0,-1){5}}
\put(140,20){\makebox(0,0){$6$}}
\put(160,40){\line(0,-1){5}}
\put(160,20){\makebox(0,0){$7$}}
\put(180,40){\line(0,-1){5}}
\put(180,20){\makebox(0,0){$8$}}
\put(200,40){\line(0,-1){5}}
\put(200,20){\makebox(0,0){$9$}}
\put(220,40){\line(0,-1){5}}
\put(220,20){\makebox(0,0){$10$}}
\put(240,40){\line(0,-1){5}}
\put(240,20){\makebox(0,0){$11$}}
\put(20,0){\makebox(0,0)[b]{$A$}}
\dottedline[.]{3}(20,40)(20,166)
\put(60,0){\makebox(0,0)[b]{$B$}}
\dottedline[.]{3}(60,40)(60,166)
\put(140,0){\makebox(0,0)[b]{$C$}}
\dottedline[.]{3}(140,40)(140,166)
\put(240,0){\makebox(0,0)[b]{$D$}}
\dottedline[.]{3}(240,40)(240,166)
\end{picture}
}

\put(0,53){\makebox(0,0)[l]{Request $B$}}

\put(75,40){\begin{picture}(260,22)(0,0)
\put(0,0){\line(1,0){260}}
\put(57,8){\fcolorbox{Black}{Black}{\makebox(0,0){\mbox{}}}}
\put(137,8){\fcolorbox{Black}{Black}{\makebox(0,0){\mbox{}}}}
\put(237,8){\fcolorbox{Black}{Black}{\makebox(0,0){\mbox{}}}}
\put(57,19){\fcolorbox{Black}{LightGray}{\makebox(0,0){\mbox{}}}}
\put(97,19){\fcolorbox{Black}{LightGray}{\makebox(0,0){\mbox{}}}}
\put(157,19){\fcolorbox{Black}{LightGray}{\makebox(0,0){\mbox{}}}}
\end{picture}
}

\put(0,100){\makebox(0,0)[l]{Request $C$}}

\put(75,87){\begin{picture}(260,22)(0,0)
\put(0,0){\line(1,0){260}}
\put(17,8){\fcolorbox{Black}{Black}{\makebox(0,0){\mbox{}}}}
\put(137,8){\fcolorbox{Black}{Black}{\makebox(0,0){\mbox{}}}}
\put(237,8){\fcolorbox{Black}{Black}{\makebox(0,0){\mbox{}}}}
\put(17,19){\fcolorbox{Black}{LightGray}{\makebox(0,0){\mbox{}}}}
\put(137,19){\fcolorbox{Black}{LightGray}{\makebox(0,0){\mbox{}}}}
\put(157,19){\fcolorbox{Black}{LightGray}{\makebox(0,0){\mbox{}}}}
\end{picture}
}

\put(0,147){\makebox(0,0)[l]{Initial}}

\put(75,134){\begin{picture}(260,22)(0,0)
\put(0,0){\line(1,0){260}}
\put(17,8){\fcolorbox{Black}{Black}{\makebox(0,0){\mbox{}}}}
\put(57,8){\fcolorbox{Black}{Black}{\makebox(0,0){\mbox{}}}}
\put(237,8){\fcolorbox{Black}{Black}{\makebox(0,0){\mbox{}}}}
\put(17,19){\fcolorbox{Black}{LightGray}{\makebox(0,0){\mbox{}}}}
\put(57,19){\fcolorbox{Black}{LightGray}{\makebox(0,0){\mbox{}}}}
\put(237,19){\fcolorbox{Black}{LightGray}{\makebox(0,0){\mbox{}}}}
\end{picture}
}

\end{picture}
\end{center}
\caption{Illustration of the 3-server example. The server positions are
given in black and the virtual positions in grey.}
\label{3-server-example}
\end{figure}
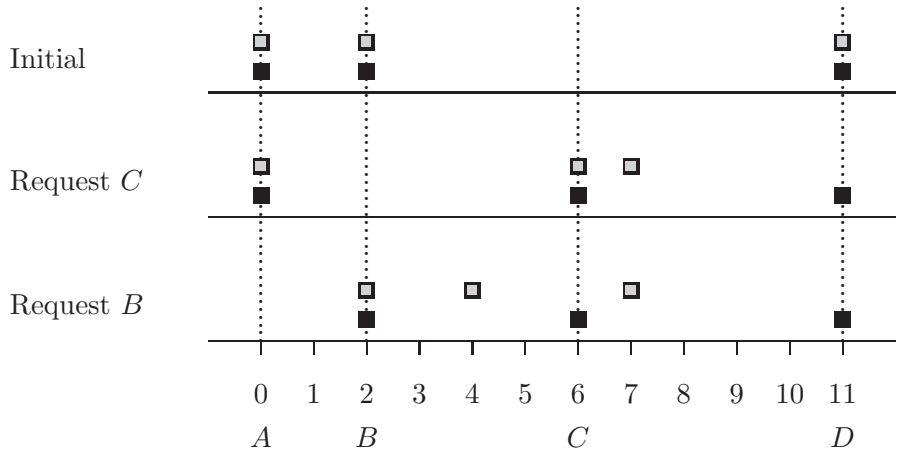

As an example, showing that some care must be taken when defining the
lazy algorithms, consider the following server problem which is slightly
more complicated than the one we consider in the rest of the paper.
We illustrate the example in Figure~\ref{3-server-example}.
There are four points
$A=0$, $B=2$, $C=6$, and $D=11$ in use,
and three servers, initially on $A$, $B$,
and $D$. We consider the request sequence $CBC$, served by \LDC.
After the first request to $C$, we have the configuration
$A$ $(A)$, $C$ $(C)$, $D$ $(7)$, where the server positions are listed
from left to right with their virtual positions in parentheses.
At the request to $B$, it becomes
$B$ $(B)$, $C$ $(4)$, $D$ $(7)$.
Now, when requesting $C$ again, note that virtually, the right-most server
is closest, but the middle server is actually on $C$.

\subsection{Quality Measures}

In analyzing algorithms for the baby server problem, we consider
input sequences $I$ of request points. An algorithm \ALG, which
treats such a sequence has some cost, which is the total distance
moved by the two servers. This cost is denoted by $\ALG(I)$.
Since $I$ is of finite length, it is clear that there exists an
offline algorithm with minimal cost. By \OPT, we refer to such an algorithm
and $\OPT(I)$ denotes the unique minimal cost of processing $I$.

All of the measures described below
can lead to a conclusion as to which one of two algorithms
is better. 
In contrast to the others, Bijective Analysis
does not quantify how much better one algorithm is than another.

\subsubsection{Competitive Analysis:}
In Competitive Analysis~\cite{G66,ST85,KMRS88}, we define
an algorithm \ALG to be $c$-competitive if there exists a constant
$\alpha$ such that for all input sequences $I$,
$\ALG(I)\leq c\,\OPT(I)+\alpha$.

\subsubsection{The Max/Max Ratio:}
The Max/Max Ratio~\cite{BDB94}
compares an algorithm's worst cost for
any sequence of length~$n$ to \OPT{}'s worst cost for any
sequence of length~$n$. 
The Max/Max Ratio of an algorithm $\ALG$,
$w_M(\ALG)$, is $M(\ALG)/M(\OPT)$, where
\[
M(\ALG)=
\limsup_{t\rightarrow\infty}\max_{\SIZE{I}=t}\ALG(I)/t.
\]

\subsubsection{The Random Order Ratio:}
Kenyon~\cite{K96} defines 
the Random Order Ratio to be
the worst ratio obtained over all sequences $I$,
comparing the expected value of an algorithm, \ALG, 
with respect to a uniform distribution of
all permutations
of $I$, to the value of \OPT{} on $I$:
\[
\limsup_{\OPT(I)\rightarrow\infty}\frac{E_{\sigma}\left[\ALG(\sigma(I))\right]}{\OPT(I)}
\]
The original context for this definition is Bin Packing for which
the optimal packing is the same, regardless of the order in which
the items are presented. Therefore, it does not make sense to
take an average over all permutations for \OPT.
For server problems, however, the order of requests in the sequence
may very well change the cost of \OPT, so we compare to \OPT's
performance, also on a random permutation of the input sequence.
In addition, taking the limit as $\OPT(I)\rightarrow\infty$, causes a
problem with analyzing \Greedy{} on the baby server problem (and
presumably other algorithms for other problems), since there is an
infinite family of sequences, $I_n$, where \OPT's cost on $I_n$ is the
same constant for all $n$, but \Greedy{}'s cost grows with $n$. Thus, we
consider the limit as the length of the sequence goes to infinity, as
in another alternative definition of the Random Order Ratio in~\cite{CCRZ08}.
 We choose to modify
the Random Order Ratio as shown to the left,
but for the results presented here, the definition to the right
would give the same:
\[
\limsup_{|I|\rightarrow\infty}\frac{E_{\sigma}\left[\ALG(\sigma(I))\right]}{E_{\sigma}\left[\OPT(\sigma(I))\right]}
\hspace{5em}
\limsup_{|I|\rightarrow\infty}  E_{\sigma}\left[ \frac{\ALG(\sigma(I))}{\OPT(\sigma(I))} \right]
\]

\subsubsection{Bijective Analysis and Average Analysis:}
In~\cite{ADLO07}, Bijective and Average Analysis are defined, as
methods of comparing two online algorithms directly. 
We adapt those definitions to the notation used here.
As with the
Max/Max Ratio and Relative Worst Order Analysis, the two algorithms
are not necessarily compared on the same sequence.

In Bijective Analysis, the sequences of a given length are mapped,
using a bijection, onto the same set of sequences. The performance
of the first algorithm on a sequence, $I$, is compared to the performance
of the second algorithm on the sequence $I$ is mapped to.
If $I_n$ denotes the set of all input sequences of length $n$, then
an online algorithm \ALG{} is no worse than an online algorithm \ALGB{}
according to Bijective Analysis if there exists an integer $n_0\geq 1$ such
that for each $n\geq n_0$, there is a bijection $f\WEHAVE I_n \rightarrow I_n$
satisfying $\ALG(I) \leq \ALGB(f(I))$ for each $I\in I_n$.
\ALG{} is strictly better than \ALGB{} if \ALG{} is no worse than
\ALGB{}, and there is no bijection showing that \ALGB{} is no worse than \ALG{}.

Average Analysis can be viewed as a relaxation of Bijective Analysis.
An online algorithm \ALG is no worse than an online algorithm \ALGB
according to Average Analysis if there exists an integer $n_0\geq 1$ such
that for each $n\geq n_0$,
$\Sigma_{I\in I_n} \ALG(I) \leq \Sigma_{I\in I_n} \ALGB(I)$.
\ALG{} is strictly better than \ALGB{} if this inequality is strict.

\subsubsection{Relative Worst Order Analysis:}

Relative Worst Order Analysis was introduced in~\cite{BF07} and extended
in~\cite{BFL07}. It compares two online algorithms directly. As with 
the Max/Max Ratio, it compares two algorithms on their worst sequence
in the same part of a partition. The partition is based on the Random
Order Ratio, so that the algorithms are compared on sequences having
the same content, but possibly in different orders.
\begin{definition}
\label{def:cost}
Let $I$ be any input sequence, and let $n$ be the length of $I$.
If $\sigma$ is a permutation on $n$ elements,
then $\sigma (I)$ denotes $I$ permuted by $\sigma$.
Let \alg{A} be any algorithm.
Then ${\mathbb A}(I)$ is the cost
 of running ${\mathbb A}$ on $I$, and
 $$\worst{\alg{A}}=\max_{\sigma}\alg{A}(\sigma(I)).$$
\end{definition}

\begin{definition}
\label{def:rwor}
For any pair of algorithms \alg{A} and \alg{B}, we define
\begin{align*}
 &\clower \; = \; \sup \left\{ c \mid \exists b \colon \forall I
 \colon \worst{\alg{A}} \geq c \, \worst{\alg{B}} - b \right\}  \text{ and }\\
 &\cupper \; = \; \inf \left\{ c \mid \exists b \colon \forall I
 \colon \worst{\alg{A}} \leq c \, \worst{\alg{B}} + b \right\} \,.
 \end{align*}
If $\clower \geq 1$ or $\cupper \leq 1$, the algorithms are said to be {\em
 comparable} and the {\em Relative Worst Order Ratio} 
 \rwor{\alg{A}}{\alg{B}} of 
 algorithm ${\alg{A}}$ to algorithm ${\alg{B}}$ is defined.
Otherwise, \rwor{\alg{A}}{\alg{B}} is undefined.
\begin{align*}
&\text{If }\clower \geq 1, \text{ then }\rwor{\alg{A}}{\alg{B}} = \cupper, \text{
 and} \\[1mm] 
& \text{if } \cupper \leq 1, \text { then }\rwor{\alg{A}}{\alg{B}} = \clower \,.
\end{align*}
If $\rwor{\alg{A}}{\alg{B}} < 1$, algorithms \alg{A} and \alg{B} are said to be
 {\em comparable in \alg{A}'s favor}.
Similarly, if $\rwor{\alg{A}}{\alg{B}} > 1$, the algorithms are said to be {\em
 comparable in \alg{B}'s favor}.

If at least one of the ratios \cupper and \cupperba is finite, then the algorithms \alg{A}
 and \alg{B} are  called {\em $(\cupper,\cupperba)$-related}.

Algorithms \alg{A} and  \alg{B} are {\em weakly comparable in \alg{A}'s favor},
1) if \alg{A} and \alg{B} are comparable in \alg{A}'s favor,
2) if \cupper is finite and \cupperba is infinite, or
3) if $\cupper \in o(\cupperba)$.
\end{definition}

An informal summary, comparing these measures is given in
Table~\ref{measures}. Note that some details are missing, including
the additive constants for asymptotic analysis.
\begin{table}[t]
\begin{center}
\begin{tabular}{|l|c|}\hline
\raisebox{0ex}[4.5ex][3ex]{\mbox{}}
Measure & Value \\ \hline\hline
\raisebox{0ex}[4.5ex][3ex]{\mbox{}} 
Competitive Ratio &
$\displaystyle CR_{\ALG} = \max_{I} \frac{\ALG(I)}{\OPT(I)}$ \\ \hline
\raisebox{0ex}[4.5ex][3ex]{\mbox{}} 
Max/Max Ratio & $\displaystyle MR_{\ALG} = \frac{\max_{|I|=n}             
\ALG(I)}{\max_{|I'|=n} \OPT({I'})}$ \\ \hline
\raisebox{0ex}[4.5ex][3ex]{\mbox{}} 
Random Order Ratio & $\displaystyle RR_{\ALG} = \max_{I}                   
\frac{E_{\sigma}\big[\ALG(\sigma(I))\big]}{E_{\sigma}\left[\OPT(\sigma(I))\right]}$ \\ \hline
\raisebox{0ex}[4.5ex][3ex]{\mbox{}} 
Relative Worst Order Ratio & $\displaystyle WR_{\ALG,\ALGB} = \max_{I}    
\frac{\max_{\sigma}\ALG(\sigma(I))}
{\max_{\sigma'}\ALGB(\sigma'(I))}$
 \\ \hline
\end{tabular}
\end{center}
\caption{Comparison of those measures which give a ratio.}
\label{measures}
\end{table}

Table~\ref{results} is a summary of the results comparing
$\LDC{}$ and $\Greedy$ on the baby server problem using each of
the measures defined. Additionally, it lists the effect of
laziness applied to \DC{}.

\begin{table}[t]
\begin{center}
\begin{tabular}{|l|l|l|}\hline
\raisebox{0ex}[3.5ex][2ex]{\mbox{}}
Measure & Favored Algorithm & \DC{} vs.\ \LDC{} \\ \hline\hline
\raisebox{0ex}[3.5ex][2ex]{\mbox{}}
Competitive Ratio & $\LDC{}$ & identical \\ \hline
\raisebox{0ex}[3.5ex][2ex]{\mbox{}} 
Max/Max Ratio & $\Greedy$ & identical \\ \hline
\raisebox{0ex}[3.5ex][2ex]{\mbox{}} 
Random Order Ratio & $\LDC{}$ & identical \\ \hline
\raisebox{0ex}[3.5ex][2ex]{\mbox{}} 
Bijective Analysis & $\Greedy{}$ & \LDC{} best \\ \hline
\raisebox{0ex}[3.5ex][2ex]{\mbox{}} 
Average Analysis & $\Greedy{}$ & \LDC{} best \\ \hline
\raisebox{0ex}[3.5ex][2ex]{\mbox{}} 
Relative Worst Order Ratio & $\LDC{}$ weakly favored
 & \LDC{} best \\ \hline
\end{tabular}
\end{center}
\caption{The second column summarizes the results comparing
$\LDC{}$ and $\Greedy$ on the baby server problem using each of the
measures defined. In addition to the information in the column,
\Greedy{} is uniquely optimal according to Bijective and Average Analysis,
and \LDC{} and \Greedy{} are $(2,\infty)$-related according to
Relative Worst Order Analysis.
The third column lists which measures distinguish between
\DC{} and its lazy variant, \LDC{}.}
\label{results}
\end{table}

\section{Competitive Analysis}
The $k$-server problem has been studied using Competitive Analysis
starting in~\cite{MMS88}.
In~\cite{CKPV91}, it is shown that on the real line, the Competitive Ratios of
\DC{} and \LDC{} are~$k$, which is optimal,
and that \Greedy{} is not competitive. The result in~\cite{MMS88},
showing that the Competitive Ratio
of \BAL{} is $n-1$ on a metric space with $n$ points if $k=n-1$,
shows that \BAL{} has the same Competitive Ratio of $2$ as \DC{}
and \LDC{} on the baby server problem.

\section{The Max/Max Ratio}
In~\cite{BDB94}, a concrete example is given with two servers and three
non-colinear points. It is observed that the Max/Max Ratio
favors the greedy algorithm over
the balance algorithm, \BAL{}.

\BAL{} behaves similarly to \LDC{} and identically on \LDC{}'s
worst case sequences.
The following theorem shows that the same conclusion is reached when
the three points are on the line.

\begin{theorem}
\label{MaxMaxGreedyBest}
\Greedy{} is better than \DC{} and \LDC{} on the baby server problem
with respect to the Max/Max Ratio, with
$\frac{w_M(\DC{})}{w_M\Greedy{})}=\frac{w_M(\LDC{})}{w_M(\Greedy{})} = 1+\frac{d-1}{d+1}$.
\end{theorem}
\begin{proof}
Given a sequence of length $n$, \Greedy{}'s maximum cost is $n$,
so $M(\Greedy{})=1$.

Since \OPT{} is at least as good as \Greedy{},
its cost is at most $n$. Thus, $M(\OPT) \leq 1$. 
To obtain a lower bound for $M(\OPT)$, we consider request sequences
consisting of repetitions of the sequence $((BA)^dC)^k$.
In each such repetition, $\OPT$ must incur a cost of at least $2d$.
Thus,
we can bound
$M(\OPT)$ by
$M(\OPT) \geq  \frac{2d}{2d+1}$.

We now determine $M(\LDC{})$, and the same argument holds for $M(\DC{})$.

For any positive integer $n$, we define the sequence
$I_n=((BA)^d BC)^pX$ of length $n$, where the length of the 
alternating $A/B$-sequence before
the $C$ is $2d+1$, $X$ is a possibly empty alternating sequence of
$A$s and $B$s starting with a $B$,
$\SIZE{X}=\MOD{n}{(2d+2)}$, and
$p=\frac{n-\SIZE{X}}{2d+2}$.

First, we claim that $I_n$ is a sequence of length $n$ where
\LDC{} has the largest average cost per move. Each move that the
right-most server, originally on $C$, makes costs $d>1$ and the
left-most server's moves cost only one. For every move the
right-most server makes from $C$ to
$B$, there are $d$ moves by the left-most server from $A$ to $B$
and thus $d$ moves back from $B$ to $A$. The subsequence
$(BA)^d$ does this with cost one for \LDC{} for every move.
Since the move after every $C$ has cost one, it is impossible to
define another sequence with a larger average cost per move.

If $\SIZE{X}<2d+1$, then the server on $C$ does not move again, and
$\LDC(s)=p(2d+2d)+\SIZE{X}=
n+\frac{(d-1)(n-\SIZE{X})}{d+1}$.

Otherwise, $\SIZE{X}=2d+1$, the server on $C$ is moved to $B$, and we obtain
$\LDC(I_n)=p(2d+2d)+\SIZE{X} +d-1=
n+\frac{(d-1)(n-\SIZE{X})}{d+1}+d-1$.

Since we are taking the supremum, we restrict our attention
to sequences where $\SIZE{X}=0$.
Thus,
$M(\LDC{})=
\frac{n+\frac{(d-1)n}{d+1}}{n}=
1+\frac{d-1}{d+1}$

Finally,
\[w_M(\Greedy{})=\frac{M(\Greedy{})}{M(\OPT{})} = \frac{1}{M(\OPT)},\]
while
\[w_M(\LDC{})=\frac{M(\LDC{})}{M(\OPT{})} =
\frac{1+\frac{d-1}{d+1}}{M(\OPT)}.\]

Since $M(\OPT)$ is bounded, $\frac{w_M(\LDC{})}{w_M(\Greedy{})} =
1+\frac{d-1}{d+1}$, which is greater than one for $d>1$.
\end{proof}

It follows from the proof of this theorem that \Greedy{} is close to optimal
with respect to the Max/Max Ratio,
since the cost of \Greedy{} divided by the cost of \OPT{} tends toward one
for large $d$.

Since \LDC{} and \DC{} perform identically on their worst sequences
of any given length, 
they also have the same Max/Max Ratio.

\section{The Random Order Ratio}

The Random Order Ratio categorizes \DC{} and \LDC as being
equally good.
The proof is structured into several lemmas below.

In the following, we use the term {\em run} to mean a sequence
of the same item in a longer sequence, and it is {\em maximal}
if it cannot be made longer by including a possible neighboring
item. For example, the three maximal runs of $A$s in
$AAABAAAABBA$ have lengths 3, 4, and 1, respectively.

The Random Order Ratio is the worst ratio obtained over all sequences,
comparing the expected value of an algorithm over all permutations
of a given sequence to the expected value of \OPT{} over all
permutations of the given sequence.
The intuition in establishing the following result is that
if one chooses a random permutation of a sequence with
many more $A$s and $B$s than $C$s, then, with high probability,
there will be sufficiently many switches between requests to $A$s and
$B$s in between each two successive occurrences of $C$s that
both \DC and \LDC will experience the full penalty compared
to OPT, i.e., after each request to $C$, they will use one server
to serve requests to both $A$ and $B$ before eventually moving
the server from $C$ to $B$, where it will stay until the next
request to $C$.

The two main components in the proof are the following:
First, even though we choose a sequence with many more $A$s and $B$s
than $C$s, we must prove that with high probability, there are
enough requests between any two $C$s.
If there are just a small constant fraction of pairs of successive $C$s
that do not have enough $A$s and $B$s in between them, we will
not get the Random Order Ratio of two that we are trying to obtain.
Second, even
though there are many requests to $A$s and $B$s in between two
consecutive $C$s, if the $A$s or $B$s, respectively, appear
as runs too frequently (many $A$s in a row, followed by many $B$s
in a row), then there will not be sufficiently many switches between
requests to $A$s and $B$s to pull a server from $C$ to $B$.
Again, we cannot afford to have this problem occur a constant fraction
of the times if we want a ratio of two.

In the proof, we choose to use $n$ requests to $A$s as well as $B$s and
$\FLOOR{\log n}$ requests to $C$s. In addition, we limit the successive requests
to $A$s and $B$s separately to $\FLOOR{\sqrt{n}}$ with high probability.
The choice of the functions $n$, $\log n$, and $\sqrt{n}$
is mostly to work with familiar functions in the lemmas below.
Many other choices of functions would work,
as long as their rates of growth are similar.
It is not quite sufficient that they are different, since we
also need to use, for instance, that $\sqrt{n}\log^2 n\in o(n)$.

We use the notation $\DOWN{n}{r}$, where $r\leq n$, for the
expression $n (n-1) (n-2) \cdots (n-r+1)$.

The following result is from~\cite{BC61}, using the index for the
last term of the summation from \cite[page~56]{BK02}.
We have substituted in our variable names:
\begin{proposition}
\label{from-statistics-article}
In a random permutation of $n$ $A$s and $n$ $B$s,
the probability that the longest run of $A$s (or $B$s) is shorter than $r$ is 
\[
\begin{array}{rcl}
P(r) & = & 1-\binom{n+1}{1}\frac{\DOWN{n}{r}}{\DOWN{2n}{r}}+
    \binom{n+1}{2}\frac{\DOWN{n}{2r}}{\DOWN{2n}{2r}}-
    \binom{n+1}{3}\frac{\DOWN{n}{3r}}{\DOWN{2n}{3r}}+
    \ldots \\[2ex]
&& + (-1)^{\FLOOR{\frac{n}{r}}} \bigbinom{n+1}{\FLOOR{\frac{n}{r}}}\frac{\DOWN{n}{\left(\FLOOR{\frac{n}{r}}r\right)}}{\DOWN{2n}{\left(\FLOOR{\frac{n}{r}}r\right)}}
\end{array}
\]
\end{proposition}

We first derive a simple lower bound on this probability.
\begin{lemma}
\label{simpler-bound-for-short-runs}
If $r\geq\log n$,
then in a random permutation of $n$ $A$s and $n$ $B$s,
the probability $P(r)$ that the longest run of $B$s
is shorter than $r$ is at least $1-\frac{n+1}{2^r}$.
\end{lemma}
\begin{proof}
We first prove that the absolute value of the terms in the expression for
$P(r)$ from Proposition~\ref{from-statistics-article} are non-increasing.
Let $1 \leq i \leq \FLOOR{\frac{n}{r}}-1$.
We consider two successive terms and show that the absolute value
of the first is at least as large as the absolute value of
the second, provided that $r\geq \log n$.
\[
\begin{array}{cl}
& \binom{n+1}{i}\frac{\DOWN{n}{ir}}{\DOWN{2n}{ir}}
  \geq
  \binom{n+1}{i+1}\frac{\DOWN{n}{(i+1)r}}{\DOWN{2n}{(i+1)r}}
\\
\Updownarrow \\
& \binom{n+1}{i}\frac{n(n-1)\cdots(n-ir+1)}{2n(2n-1)\cdots(2n-ir+1)}
  \geq
  \binom{n+1}{i+1}\frac{n(n-1)\cdots(n-(i+1)r+1)}{2n(2n-1)\cdots(2n-(i+1)r+1)}
\\
\Updownarrow \\
& \binom{n+1}{i}
  \geq
  \binom{n+1}{i+1}\frac{(n-ir)(n-ir-1)\cdots(n-(i+1)r+1)}{
                        (2n-ir)(2n-ir-1)\cdots(2n-(i+1)r+1)} \\
\Uparrow \\
& \frac{(n+1)!}{i!(n+1-i)!}
  \geq
  \frac{(n+1)!}{(i+1)!(n-i)!}\left(\frac{n-ir}{2n-ir}
                             \right)^{ir} \\
\Uparrow \\
& 1 \geq \frac{n-i+1}{i+1}\left(\frac{1}{2}\right)^{ir} \\
\Updownarrow \\
& 2^{ir} \geq \frac{n-i+1}{i+1}\\
\Uparrow \\
& r \geq \log n
\end{array}
\]
where the first implication follows from considering the fractions
of corresponding factors from the numerator and denominator and
choosing the largest.

Having shown that the terms are non-increasing, it follows that
$P(r) \geq 1-\binom{n+1}{1}\frac{\DOWN{n}{r}}{\DOWN{2n}{r}}$,
i.e., dropping all but the first two terms.
Since, for corresponding factors in $\DOWN{n}{r}$ and $\DOWN{2n}{r}$,
we have that $\frac{n-j}{2n-j}\leq\frac{1}{2}$, we can conclude that
$P(r)\geq 1-\frac{n+1}{2^r}$.
\end{proof}

We can use this lemma to show that switches between $A$s and $B$s
occur quite often.
\begin{lemma}
\label{enough-switches}
Let $I_n=A^n B^n$.
For any $\varepsilon>0$, there exists an $n_0$ such that
for all $n\geq n_0$, the probability when selecting a random
permutation of $I_n$ that all maximal
runs of $A$s (or $B$s)
have lengths at most $\FLOOR{\sqrt{n}}$
is at least $1-\varepsilon$.
\end{lemma}
\begin{proof}
By Lemma~\ref{simpler-bound-for-short-runs}, for any given $n$,
the probability is at least
$1-\frac{n+1}{2^{\FLOOR{\sqrt{n}}}}$.
Since $n+1\in o(2^{\sqrt{n}})$, this probability approaches one
for increasing values of $n$.
\end{proof}

Now we show that when having so few $C$s compared to $A$s and $B$s,
we can be almost certain to find a large number of $A$s and $B$s
between two successive occurrences of $C$s.
\begin{lemma}
\label{long-enough-runs}
For any $\varepsilon>0$, there exists an $n_0$ such that
for all $n\geq n_0$, the probability when selecting a random
permutation of $I_n=A^n B^n C^{\FLOOR{\log n}}$ that all maximal runs
of $A$s and $B$s (looking at $A$s and $B$s as the same item)
have length at least $(2d+2)\FLOOR{\sqrt{n}}$
is at least $1-\varepsilon$.
\end{lemma}
\begin{proof}
We do not distinguish between $A$s and $B$s here, so we just use that
there are a total of $2n$ of them, and refer to all of them as $X$s.

To compute the probability, we consider the number of ways the $C$s
can be placed as dividers into a sequence of $2n$ $X$s, creating
$\FLOOR{\log n} + 1$ groups. The standard
method is to consider $2n+\FLOOR{\log n}$ positions and place the $C$s
in $\FLOOR{\log n}$ of these, which can be done in $\binom{2n + \FLOOR{\log n}}{\FLOOR{\log n}}$ ways.
Similarly, if we want $(2d+2)\FLOOR{\sqrt{n}}$ $X$s in each group,
we may reserve these $(2d+2)\FLOOR{\sqrt{n}}(\FLOOR{\log n} + 1)$ $X$s
and just consider the division of the remaining $X$s. Thus, this can be
done in $\bigbinom{2n - (2d+2)\FLOOR{\sqrt{n}}(\FLOOR{\log n} + 1) + \FLOOR{\log n}}{\FLOOR{\log n}}$ ways.

We now find a lower bound on the probability of there being
this many $A$s and $B$s between $C$s using the above counting argument:

\[
\begin{array}{cl}
  & \displaystyle
    \frac{\bigbinom{2n - (2d+2)\FLOOR{\sqrt{n}}(\FLOOR{\log n} + 1) + \FLOOR{\log n}}{\FLOOR{\log n}}}{
          \bigbinom{2n + \FLOOR{\log n}}{\FLOOR{\log n}}}  \\[6ex]
= & \displaystyle
    \frac{\DOWN{(2n - (2d+2)\FLOOR{\sqrt{n}}(\FLOOR{\log n} + 1) + \FLOOR{\log n})}{\FLOOR{\log n}}}{
          \DOWN{(2n + \FLOOR{\log n})}{\FLOOR{\log n}}} \\[3ex]
\geq & \displaystyle
       \left(\frac{2n - (2d+2)\FLOOR{\sqrt{n}}(\FLOOR{\log n} + 1) + 1}{
                   2n+1}\right)^{\FLOOR{\log n}} \\[3ex]
= & \displaystyle
    \left(1-\frac{(2d+2)\FLOOR{\sqrt{n}}(\FLOOR{\log n} + 1)}{2n+1}\right)^{\FLOOR{\log n}} \\
\end{array}
\]
where the inequality follows from considering corresponding factors
in the numerator and denominator, and using the smallest fraction
of these.

Using the binomial theorem, this last expression can be written
\[
\begin{array}{cl}
&
\displaystyle
\sum_{i=0}^{\FLOOR{\log n}}\binom{\FLOOR{\log n}}{i}\left(
            \frac{-(2d+2)\FLOOR{\sqrt{n}}(\FLOOR{\log n} + 1)}{2n+1}
                                    \right)^i
\\[4ex]
= &
\displaystyle
1 - \FLOOR{\log n} \left(\frac{(2d+2)\FLOOR{\sqrt{n}}(\FLOOR{\log n} + 1)}{2n+1}\right) + T
\end{array}
\]
where $T$ contains the additional terms of the binomial expansion.

We now argue that the absolute values of successive terms in $T$ decrease
for large enough $n$:
\[
\begin{array}{cl}
& \binom{\FLOOR{\log n}}{i}\left(\frac{(2d+2)\FLOOR{\sqrt{n}}(\FLOOR{\log n} + 1)}{2n+1}\right)^i
>
\binom{\FLOOR{\log n}}{i+1}\left(\frac{(2d+2)\FLOOR{\sqrt{n}}(\FLOOR{\log n} + 1)}{2n+1}\right)^{i+1}
\\
\Updownarrow \\
& \frac{\DOWN{\FLOOR{\log n}}{i}}{i!} > \frac{\DOWN{\FLOOR{\log n}}{i+1}\,(2d+2)\FLOOR{\sqrt{n}}(\FLOOR{\log n} + 1)}{(i+1)!(2n+1)}\\
\Updownarrow \\
& 1 > \frac{(\FLOOR{\log n} - i)(2d+2)\FLOOR{\sqrt{n}}(\FLOOR{\log n} + 1)}{(i+1)(2n+1)}
\end{array}
\]
Since $\sqrt{n}\log^2 n \in o(n)$,
this holds when $n$ is sufficiently large.

For $n$ large enough, this means that $T\geq 0$ and
the probability we are computing will be bounded from below by
$1 - \FLOOR{\log n} \left(\frac{(2d+2)\FLOOR{\sqrt{n}}(\FLOOR{\log n} + 1)}{2n+1}\right)$.

Again, since $\sqrt{n}\log^2 n \in o(n)$,
the probability approaches one as $n$ increases.
\end{proof}

With the use of the lemmas above, we can establish the theorem.
\begin{theorem}
\label{random-order-lazy}
\DC and \LDC both have the Random Order Ratio two.
\end{theorem}
\begin{proof}
The upper bounds follow directly from the fact that their Competitive
Ratios are two. Thus, if that is the factor on worst case sequences,
clearly the expected ratio cannot be worse, since the averages for these 
algorithms and \OPT{} is over the same set of sequences.

For the lower, let $I_n=A^n B^n C^{\FLOOR{\log n}}$.
We show that for any $\varepsilon>0$, there exists an $n_0$
so that for $n\geq n_0$,
the probability of \DC and \LDC incurring a cost of a factor two more than
\OPT is at least $1-\varepsilon$. This immediately implies
that the expected cost of the two algorithms cannot be smaller,
giving us the result.

By Lemma~\ref{long-enough-runs}, there exists an $n'$ so that for
all $n\geq n'$, the probability that all maximal runs
of $A$s and $B$s have length at least $(2d+2)\FLOOR{\sqrt{n}}$
is at least $1-\frac{\varepsilon}{2}$.

Considering only the $A$s and $B$s, by Lemma~\ref{enough-switches},
there exists an $n''$ so that for all $n\geq n''$,
the probability that all maximal runs of $A$s and $B$s, respectively,
have lengths at most $\FLOOR{\sqrt{n}}$ is at least $1-\frac{\varepsilon}{2}$.

Thus, for all $n\geq\max\SET{n',n''}$, the probability of having
both properties is at least $1-\varepsilon$,
and we argue that in this case, the cost
of \DC and \LDC are a factor two
larger than the cost of \OPT.

Since the number of $A$s and $B$s between two $C$s is at least
$(2d+2)\FLOOR{\sqrt{n}}$ and the length of maximal runs of
$A$s and $B$s, respectively, is at most $\FLOOR{\sqrt{n}}$,
there must at least $2d+2$ runs in between two successive $C$s,
and at least $2d+1$ runs if we want to count from 
the first run of $B$s.

For both algorithms, this is sufficient for the algorithm to move
the server from $C$ to $B$. \DC will have both servers on $B$
after the $d$th run of $B$s has been processed,
whereas for \LDC, the right-most server will only virtually
be at $B$ at that point, but will be moved there at the $(d+1)st$
run of $B$s.

For each $C$, \OPT incurs the cost $2d$ of moving a server from
$C$ to $B$ and back again, and it incurs cost $d$ after the last $C$. 
The online algorithms have the same
cost, plus the additional cost of moving a server back and forth
between $A$ and $B$ until the server from $C$ is moved to $B$.
This additional cost consists of $2d$ complete moves
from $A$ to $B$ and back.

Asymptotically, the requests after the last $C$
can be ignored, so this gives the ratio $4d/2d=2$.
\end{proof}

This result, saying that \LDC{} and
\DC{} are equivalent according the Random Order Ratio, is
an example of where a counter-intuitive result is clearly due to the
intermediate comparison to \OPT, because on some of the sequences
where \LDC{} and \DC{} do worst compared to \OPT, they do 
equally badly compared to \OPT. We illustrate this problem with
the intermediate comparison to \OPT{} by showing below how avoiding this
comparison could give the result that \LDC{} is
better than \DC.

If the definition was modified in the most straightforward
manner to allow direct comparison of algorithms, one would first
note that for any sequence $I$,
$E_{\sigma}[\DC(\sigma(I))]\geq E_{\sigma}[\LDC(\sigma(I))]$,
by the laziness observation. Then, one would consider some
families of sequences with relatively large numbers of $C$s and show that
\LDC's cost is some constant fraction better than \DC's on
random permutations of that sequence. 

For example,
let $I= (CABC)^n$. Whenever the subsequence $CABC$ occurs in
$\sigma(I)$, \DC{} moves a server from $C$ towards $B$ and back
again, while moving the other server from $A$ to $B$. In contrast,
\LDC{} lets the server on $C$ stay there, and has cost two less than
\DC{}.

One can show that
the expected number of occurrences of $CABC$ in $\sigma(I)$
is at least $\frac{n}{16}$ (any constant fraction of $n$ would
illustrate the point)
by considering any of the possible starting locations
for this pattern, $1\leq i\leq 4n-3$, and noting that the probability that
the pattern $CABC$ begins there is $\frac{1}{2}\cdot\frac{n}{4n-1}
\cdot\frac{n}{4n-2}\cdot\frac{2n-1}{4n-3}$.
By the linearity of expectations, the expected
number of occurrences of $CABC$ is $(\frac{1}{2}\cdot\frac{n}{4n-1}
\cdot\frac{n}{4n-2}\cdot\frac{2n-1}{4n-3})\cdot (4n-3)=
\frac{1}{2}\cdot\frac{1}{2}\cdot\frac{n^2}{4n-1}\geq \frac{n}{16}$.

The expected costs of both \OPT{} and
\LDC{} on $\sigma(I)$ are also bounded above and below by some 
constants times $n$. Thus, \LDC's ``modified random order ratio'' will be
less than \DC's.

It is easier to compare \Greedy{} and \LDC{} using the
(original) Random Order
Ratio, getting a result very similar to that of Competitive Analysis:
\LDC{} is strictly better than \Greedy{}.

\begin{theorem}
\DC{} and \LDC{} are better than \Greedy{} on the baby server problem
with regards to the Random Order Ratio.
\end{theorem}
\begin{proof}
As noted in the proof of Theorem~\ref{random-order-lazy},
since the Competitive Ratios of both \DC{} and \LDC{} are two, 
their Random Order Ratios are also at most two.

Consider all permutations of the sequence $I_n=(BA)^{\frac{n}{2}}$.
We consider positions from 1 through $n$ in these sequences.
We again
refer to a maximal consecutive subsequence consisting entirely of either
$A$s or $B$s as a {\em  maximal run}.

Given a sequence containing $h$ $A$s and $t$ $B$s, one can see
from well known results that
the expected number of maximal runs is $1+\frac{2ht}{h+t}$:
In \cite[Problem~28, Chapter~9, Page~240]{F68}, it is
stated that the expected number of runs of $A$s is $\frac{h(t+1)}{h+t}$, so
the expected number of runs of $B$s is $\frac{t(h+1)}{h+t}$. One can
see that this holds for $A$s by considering the probability that a
run of $A$s starts at some index $i$ in the sequence. The probability
that it starts at the beginning of the sequence, at index $i=1$,
is the probability that the first element is an $A$, $\frac{h}{h+t}$.
The probability that it starts at some index $i>1$ is the probability that
there is a $B$ at index $i-1$ and an $A$ at index $i$, $\frac{t}{h+t}\cdot
\frac{h}{h+t-1}$. By the linearity of expectations, the expected number
of runs of $A$s is thus 
$\frac{h}{h+t}+\sum_{i=2}^{h+t}\frac{th}{(h+t)(h+t-1)}=\frac{h(t+1)}{h+t}$. 
Adding the 
expectations for $A$s and $B$s gives the result $1+\frac{2ht}{h+t}$.
Thus, with $h=t=\frac{n}{2}$, we get that $\frac{n}{2} + 1$ is the
expected number of runs.

The cost of \Greedy{} is equal to the number of runs if the first run
is a run of $B$s. Otherwise, the cost is one smaller.
Thus, \Greedy{}'s expected cost on a permutation of $I_n$ is
$\frac{n}{2} + \frac{1}{2}$.

The cost of \OPT{} for any permutation of $I_n$ is $d$,
since it simply moves the server from $C$ to $B$ on the
first request to $B$ and has no other cost after that.

Thus, the Random Order Ratio is $\frac{\frac{n}{2}+\frac{1}{2}}{d}$, which,
as $n$ tends to infinity, is unbounded.
\end{proof}

The same argument shows that \BAL{} is better than \Greedy{}
with respect to the Random Order Ratio.

\section{Bijective Analysis}

Bijective analysis correctly distinguishes between \DC{} and
\LDC{}, indicating that the latter is the better algorithm.
This follows from the following general theorem about lazy
algorithms, and the fact that there are some sequences where
one of \DC's servers repeatedly moves from $C$ towards $B$, but
moves back to $C$ before ever reaching $B$, while \LDC's server
stays on $C$.

\begin{theorem}
\label{lazybijective}
The lazy version of any algorithm for the baby
server problem is at least as good as the original algorithm
according to both Bijective Analysis and Average Analysis. 
\end{theorem}
\begin{proof}
By the laziness observation, the identity function, $id$, is
a bijection such that $\LALG(I)\leq \ALG(id(I))$ for all
sequences $I$. If an algorithm is better than another algorithm with 
regards to Bijective
Analysis, then it is also better with regards to Average
Analysis~\cite{ADLO07}.
\end{proof}

We first show that \Greedy{} is at least as 
good as any other lazy algorithm; including \LDC{} and \BAL{}.

\begin{theorem}
\label{Greedy-best}
\Greedy{} is at least as good as any other
lazy algorithm \LAZY{}
for the baby server problem according to Bijective Analysis.
\end{theorem}
\begin{proof}
Since \Greedy{} has cost zero for the sequences consisting of only the
point $A$ or only the point $C$ and cost one for the point $B$,
it is easy to define a bijection $f$ for sequences of length one,
such that $\Greedy(I)\leq \LAZY(f(I))$.
Suppose that for all sequences of length~$k$
we have a bijection, $f$, from \Greedy{}'s sequences to
$\LAZY$'s sequences, such that for
each sequence $I$ of length~$k$, $\Greedy(I)\leq \LAZY(f(I))$.
To extend this to length~$k+1$, consider the three
sequences formed from a sequence $I$ of length~$k$ by adding one of the three
requests $A$, $B$, or~$C$ to the end of $I$, and the three sequences
formed from $f(I)$ by adding each of these points to the end of $f(I)$.
At the end of sequence $I$, \Greedy{} has its two servers
on different points, so two of these new sequences
have the same cost for \Greedy{} as on $I$ and one has cost
exactly 1 more. Similarly, $\LAZY$ has its two servers
on different points at the end of $f(I)$, so two of
these new sequences have the same cost for $\LAZY$ as
on $f(I)$ and one has cost either 1 or $d$ more. This
immediately defines a bijection $f'$ for sequences of length
$k+1$ where $\Greedy(I)\leq \LAZY(f'(I))$ for all $I$ of length~$k+1$.
\end{proof}

\begin{corollary}
\label{bijective-greedy-optimal}
\Greedy{} is the unique optimal algorithm with regards to Bijective and
Average Analysis.
\end{corollary}
\begin{proof}
Note that the proof of Theorem~\ref{Greedy-best} shows that
\Greedy{} is strictly better than any lazy algorithm which ever
moves the server away from $C$, so it is better than any other lazy
algorithm with regards to Bijective Analysis.
By Theorem~\ref{lazybijective}, it is better than any
algorithm. Again, since separations with respect to Bijective Analysis
also hold for Average Analysis, the result also
holds for Average Analysis.
\end{proof}

According to Bijective Analysis, there is also a
unique worst algorithm among compliant
server algorithms for the baby server problem:
If $p$ is in between the two servers, the algorithm moves the 
server that is furthest
away to the request point. If $p$ is on the same
side of both servers, the nearest server moves to $p$.
Again, due to the problem formulation, ties cannot occur
(and the server on $A$ is never moved).
The proof that this algorithm is unique worst
is similar to the proof of Theorem~\ref{Greedy-best}, but now with 
cost $d$ for every actual move.

\begin{lemma}
\label{LDC-speed-bijection}
If $a\leq b$,
then there exists a bijection
\[\sigma_n\WEHAVE\{ A,B,C\}^n\rightarrow \{ A,B,C\}^n\]
such that $\aLDC(I)\leq \bLDC(\sigma_n(I))$ for all sequences $I \in
\{ A,B,C\}^n$.
\end{lemma}
\begin{proof}
We use the bijection from the proof of Theorem~\ref{Greedy-best},
showing that \Greedy{} is the unique best algorithm,
but specify the bijection completely, as opposed to allowing some
freedom in deciding the mapping in the cases where we are extending
by a request where the algorithms already have a server.
Suppose that the bijection $\sigma_n$ is already defined.
Consider a sequence $I_n$ of length $n$ and the three possible ways,
$I_nA$, $I_nB$ and $I_nC$, of extending it to length $n+1$. Suppose
that \aLDC{} has servers on points $X_a,Y_a\in \{ A,B,C\}$ after
handling the sequence $I_n$, and \bLDC{}
has servers on points $X_b,Y_b\in \{ A,B,C\}$ after handling $\sigma_n(I_n)$.
Let $Z_a$ be the point where \aLDC{} does not have a server and
$Z_b$ the point where \bLDC{} does not. Then $\sigma_{n+1}(I_nZ_a)$ is
defined to be $\sigma_n(I_n)Z_b$. In addition, since the algorithms
are lazy, both algorithms have their servers on two different points
of the three possible, so there must be at least one point $P$
where both algorithms have a server.
Let $U_a$ be the point in $\{ X_a,Y_a\}\setminus \{ P\}$ and
$U_b$ be the point in $\{ X_b,Y_b\}\setminus \{ P\}$. Then,
$\sigma_{n+1}(I_nP)$ is defined to be $\sigma_n(I_n)P$ and
$\sigma_{n+1}(I_nU_a)$ to be $\sigma_n(I_n)U_b$.

Consider running \aLDC{} on a sequence $I_n$ and \bLDC{} on $\sigma_n(I_n)$
simultaneously. The sequences are clearly constructed so that,
at any point during this simultaneous execution, both algorithms
have servers moving or neither does.

The result follows if we can show that \bLDC{} moves
away from and back to $C$ at least as often as \aLDC{} does.
By construction, the two sequences, $I_n$ and $\sigma_n(I_n)$,
will be identical up to the point where \bLDC{} (and possibly \aLDC{})
moves away from $C$ for the first time. In the remaining part
of the proof, we argue that if \aLDC{} moves away from and back
to $C$, then \bLDC{} will also do so before \aLDC{} can do it again.
Thus, the total cost of \bLDC{} will be at least that of \aLDC{}.

Consider a request causing the slower algorithm, \aLDC{},
to move a server away from $C$.

If \bLDC{} also moves a server away from $C$ at this point,
both algorithms have their servers on $A$ and $B$, and the two sequences
continue identically until the faster algorithm again moves a
server away from $C$
(before or at the same time as the slower algorithm does).

If \bLDC{} does not move a server away from $C$ at this point, since,
by construction, it does make a move, it moves a server from $A$ to $B$.
Thus, the next time both algorithms move a server,
\aLDC{} moves from $B$ to $C$ and
\bLDC{} moves from $B$ to $A$. Then both algorithms have servers
on $A$ and $C$.
Since \aLDC{} has just moved a server to $C$, whereas \bLDC{}
must have made at least one move from $A$ to $B$ since it placed a
server at $C$, \bLDC{} must, as the faster algorithm, make
its next move away from $C$ strictly before \aLDC{} does so.
In conclusion, the sequences will be identical until the
faster algorithm, \bLDC{}, moves a server away from $C$.
\end{proof}

\begin{theorem}
According to Bijective Analysis and Average Analysis, 
slower variants of \LDC{} are better than faster variants
for the baby server problem.
\end{theorem}
\begin{proof}
Follows immediately from Lemma~\ref{LDC-speed-bijection}
and the definition of the measures.
\end{proof}

Thus, the closer a variant of \LDC{} is to \Greedy{}, the better Bijective
and Average Analysis predict that it is.

\section{Relative Worst Order Analysis}
Similarly to the Random Order Ratio and
Bijective Analysis, Relative Worst Order Analysis
correctly distinguishes between \DC{} and
\LDC{}, indicating that the latter is the better algorithm.
This follows from the following general theorem about lazy
algorithms, and the fact that there are some sequences where
one of \DC's servers repeatedly moves from $C$ towards $B$, but
moves back to $C$ before ever reaching $B$, while \LDC's server
stays on $C$. 

Let $I_{\ALG}$ denote a worst ordering of the sequence $I$
for the algorithm \ALG.

\begin{theorem}
\label{RWOR-laziness}
The lazy version of any algorithm for the baby server problem is
at least as good as the original algorithm
according to Relative Worst Order Analysis.
\end{theorem}
\begin{proof}
By the laziness observation, for any request sequence $I$,
$\LALG(I_{\LALG})\leq \ALG(I_{\LALG}) \leq \ALG(I_{\ALG})$.
\end{proof}

\begin{theorem}
\DC{} (\LDC{}) and \Greedy{} are $(2,\infty)$-related and are thus
weakly comparable in \DC{}'s (\LDC{}'s) favor for the baby server problem
according to Relative Worst Order Analysis.
\end{theorem}
\begin{proof}
We write this proof for \DC{}, but exactly the same holds for \LDC{}.
First we show that $c_{\text{u}}(\Greedy,\DC)$ is unbounded.
Consider the sequence $(BA)^{\frac{n}{2}}$.
As $n$ tends to infinity, \Greedy{}'s cost is unbounded, whereas
\DC{}'s cost is at most $3d$ for any permutation.

Next we turn to $c_{\text{u}}(\DC,\Greedy)$.
Since the Competitive Ratio of \DC{} is $2$,
for any sequence $I$ and some constant $b$,
$\DC(I_{\DC})\leq 2\OPT(I_{\DC})+b
             \leq 2\Greedy(I_{\DC})+ b
             \leq 2\Greedy(I_{\Greedy})+b$.
Thus, $c_{\text{u}}(\DC,\Greedy)\leq 2$.

For the lower bound, consider a family of sequences
\[I_p=(BABA...BC)^p,\]
where the length of the alternating $A/B$-sequence before
the $C$ is $2d+1$.

$\DC(I_p)=p(4d)$.

A worst ordering for \Greedy{} alternates $A$s and $B$s. Since there is
no cost for the $C$s and the $A/B$ sequences start and end with $B$s,
$\Greedy(\sigma(I_p)) \leq p(2d) +1$ for any permutation $\sigma$.

Then, $c_{\text{u}}(\DC,\Greedy)\geq \frac{p(4d)}{p(2d) +1}$.
As $p$ goes to infinity, this approaches $2$.

Thus, \DC{} and \Greedy{} are weakly comparable in \DC{}'s favor.
\end{proof}

Recall in the following that for clarity in the exposition,
we assume that $a$ divides $d$.
By the definition of \aLDC{},
a request for $B$ is served by the right-most server
if it is within a virtual distance of no more than $a$ from $B$ and
the other server is at $A$.
Thus, when the left-most server moves and its virtual move
is over a distance of $l$, then the right-most server
virtually moves a distance $al$.
When the right-most server moves and its virtual move
is over a distance of $al$, then the left-most server
virtually moves a distance of $l$.

In the results that follow, we frequently look at the worst
ordering of an arbitrary sequence.
\begin{definition}
The {\em canonical worst ordering} of a sequence, $I$,
for an algorithm \ALG{} is
the sequence produced by 
allowing the cruel adversary
(the one which always lets the next request be the unique point
where \ALG{} does not currently have a server) to choose requests
from the multiset defined from $I$. This process continues until
there are no requests remaining in the multiset for the point where
\ALG{} does not have a server.
The remaining points from the multiset are
concatenated to the end of this new request sequence in any order.
\end{definition}

The canonical worst ordering of a sequence for \aLDC{} is as follows.

\begin{proposition}
Consider an arbitrary sequence $I$ containing 
$n_A$ $A$s, $n_B$ $B$s, and $n_C$ $C$s. A
canonical worst ordering of $I$ for \aLDC{} is
\[I_a=(BABA...BC)^{p_a}X,\]
where the length of the alternating $A/B$-sequence before
the $C$ is $2\DAF+1$
(recall that we assume that $\DAF$ is integral).
Here, $X$ is a possibly empty sequence. The first part of $X$ is an
alternating sequence of $A$s and $B$s, starting with a $B$, until
there are not both $A$s and $B$s left. Then we continue with all
remaining $A$s or $B$s, followed by all remaining $C$s.
Finally,
\[p_a=\MIN{\FLOOR{\frac{n_A}{\DAF}},
        \FLOOR{\frac{n_B}{\DAF+1}},
        n_C}.\]
\end{proposition}

\begin{lemma}
\label{a-LDC-worst}
Let $I_a$ be the canonical worst ordering of $I$ for \aLDC{}.
$I_a$ is a worst permutation of $I$ for \aLDC{}, and
the cost for \aLDC{} on $I_a$ is $c_a$, where $p_a(2\DAF +2d)\leq
c_a\leq p_a(2\DAF +2d)+2\DAF +d$.
\end{lemma}
\begin{proof}
Consider a request sequence,
$I$. Between any two moves from $B$ to $C$, there must have been
a move from $C$ to $B$. Consider one such move. Between the
last request to $C$ and this move, the other server must move
from $A$ to $B$ exactly $\DAF$ times, which requires some first
request to $B$ in this subsequence, followed by at least $\DAF$ occurrences
of requests to $A$, each followed by a request to $B$, the last one causing
the move from $C$ to $B$. (Clearly, extra
requests to $A$ or $B$ could also occur, either causing moves or
not.) Thus, for every move from $B$ to $C$, there must be at
least $\DAF+1$ $B$s, $\DAF$ $A$s and one $C$. Thus, the number of
moves from $B$ to $C$ is bounded from above by $p_a$.
There can be at most one more move from $C$ to $B$ than from $B$ to
$C$. If such a move occurs, there are no more $C$s after that in the
sequence. Therefore, the sequences defined above give the maximal
number of moves of distance $d$ possible. More $A$s or $B$s in any
alternating $A/B$-sequence would not cause additional moves (of either
distance one or $d$), since each extra
point requested would already have a server. Fewer $A$s or $B$s
between two $C$s would
eliminate the move away from $C$ before it was requested again.
Thus, the canonical worst ordering is a worst ordering of $I$.

Within each of the $p_a$ repetitions of $(BABA...BC)$, each of the requests
for $A$ and all but the last request for $B$ cause a move of distance one,
and the last two requests each cause a move of distance $d$, giving the
lower bound on $c_a$. Within $X$, each of the first $2\DAF$ requests 
could possibly cause a move of distance one, and this could be followed 
by a move of distance $d$. After that, no more moves occur. Thus, adding
costs to the lower bound gives the upper bound on $c_a$.
\end{proof}

\begin{theorem}
\label{a-b-LDC-related}
If $a\leq b$, then \aLDC{} and \bLDC{} are
$\left(\frac{1+\frac{1}{a}}{1+\frac{1}{b}}, \frac{b+1}{a+1}\right)$-related
for the baby server problem
according to Relative Worst Order Analysis.
\end{theorem}
\begin{proof}
By Lemma~\ref{a-LDC-worst},
in considering \aLDC's performance in comparison with \bLDC's, the
asymptotic ratio depends only on the values $p_a$ and $p_b$ defined 
for the canonical worst orderings $I_a$ and $I_b$
for \aLDC and \bLDC, respectively. Since $a\leq b$, the
largest value of $\frac{p_a}{p_b}$ occurs when $p_a=n_C$, since more
$C$s would allow more moves of distance $d$ by \bLDC{}. Since the
contribution of $X$ to \aLDC's cost can be considered to be a constant,
we may assume that 
$n_A=n_C\DAF$ and $n_B=n_C\left(\DAF+1\right)$.

When considering \bLDC{}'s canonical worst ordering of this sequence,
all the $C$s will be used in the initial part.
By Lemma~\ref{a-LDC-worst}, we obtain the following ratio, for some
constant $c$:
\[
\frac{(2\DAF+2d)n_C}{(2\DBF+2d)n_C+c}
=
\frac{(\frac{1}{a}+1)n_C}{(\frac{1}{b}+1)n_C+\frac{c}{2d}}
\]

Similarly, a  sequence giving the largest value of
$\frac{p_b}{p_a}$ will have $p_b= \FLOOR{\frac{n_A}{\frac{d}{b}}}$, since
more $A$s would allow \aLDC{} to have a larger $p_a$. Since
the contribution of $X$ to \bLDC{} can be considered to be a
constant, we may assume that
$n_A=n_C\DBF$, $n_B=n_C\left(\DBF+1\right)$, and $p_b = n_C$.

Now, when considering \aLDC{}'s worst permutation of this sequence,
the number of periods, $p_a$, is restricted by the number of $A$s.
Since each period has $\DAF$ $A$s,
$p_a=\FLOOR{\frac{n_A}{\DAF}}=
\FLOOR{\frac{n_C\DBF}{\DAF}}$.
After this, there are a constant number of $A$s remaining,
giving rise to a constant additional cost $c'$.

Thus, the ratio is the following:
\[
\frac{(2\DBF+2d)n_C}{(2\DAF+2d)\FLOOR{n_C\frac{a}{b}}+c'}
=
\frac{(\frac{1}{b}+1)n_C}{(\frac{1}{a}+1)\FLOOR{n_C\frac{a}{b}}+\frac{c'}{2d}}
=
\frac{(1+b)n_C}{(1+a)n_C + c''},
\]
for some constant $c''$.
Considering the two ratios relating \bLDC{}'s and \aLDC{}'s worst
permutations asymptotically
as $n_C$ goes to infinity, we obtain that \aLDC{} and \bLDC{} are
$\left(\frac{1+\frac{1}{a}}{1+\frac{1}{b}}, \frac{b+1}{a+1}\right)$-related.
\end{proof}

Although with the original definition of relatedness in Relative Worst Order
Analysis, the values are not interpreted further, one could use the
concept of
{\em better performance}~(see~\cite{DLOM09})
from Relative Interval Analysis to compare two algorithms using Relative
Worst Order Analysis.
Using the previous result, we show that \LDC{} has better performance 
than \bLDC{}
for $b\not=1$.
Again, for clarity, we consider integral cases in the following result.
\begin{theorem}
\label{LDC-compared-bLDC}
Consider the baby server problem evaluated according to Relative Worst Order Analysis.
For $b>1$ such that $\frac{d}{b}$ is integral, \LDC{} and \bLDC{}
are $(r,r_b)$-related for some $r$ and $r_b$ where $1<r<r_b$.
For $a<1$ such that $\frac{1}{a}$ is integral, \aLDC{} and \LDC{} 
are $(r_a,r)$-related for some $r_a$ and $r$ where $1<r<r_a$.
\end{theorem}
\begin{proof}
By Theorem~\ref{a-b-LDC-related}, \aLDC{} and \bLDC{} are
$\left(\frac{1+\frac{1}{a}}{1+\frac{1}{b}}, \frac{b+1}{a+1}\right)$-related.

To see that $\frac{1+\frac{1}{a}}{1+\frac{1}{b}} < \frac{b+1}{a+1}$ when
$1=a<b$, note that this holds if and only if $(1+\frac{1}{a})(a+1)=4<
(1+\frac{1}{b})(b+1)$, which clearly holds for $b>1$.
Hence, if \LDC{} and \bLDC{} are $(c_1,c_2)$-related,
then $c_1<c_2$.

To see that $\frac{1+\frac{1}{a}}{1+\frac{1}{b}} > \frac{b+1}{a+1}$ when
$a<b=1$, note that this holds if and only if $(1+\frac{1}{a})(a+1)>4=
(1+\frac{1}{b})(b+1)$. This clearly holds for $a<1$.
Thus, \aLDC{} and \LDC{} are $(c_1,c_2)$-related, where $c_1>c_2$.
\end{proof}

The algorithms \aLDC{} and \iaLDC{}
are in some sense of equal quality:

\begin{corollary}
If $\frac{1}{a}$ and $\frac{d}{b}$ are integral and $b=\frac{1}{a}$,
then \aLDC{} and \bLDC{} are $(b,b)$-related
\end{corollary}

Theorem~\ref{LDC-compared-bLDC} shows that \LDC{} is in some sense
optimal among the \aLDC{} algorithms.
We now set out to prove that \LDC{} is an optimal algorithm
in the following sense:
there is no other algorithm \ALG{} such that \LDC{} and \ALG{} are comparable
and \ALG{} is strictly better or
such that \LDC{} and \ALG{} are weakly comparable in \ALG{}'s favor.

\begin{theorem}
\label{LDC-is-optimal}
\LDC{} is optimal
for the baby server problem
according to Relative Worst Order Analysis.
\end{theorem}
\begin{proof}
In order for \LDC{} and \ALG{} to be comparable in \ALG{}'s favor, \ALG{}
has to be comparable to \LDC{}
and perform more than an additive constant better on some infinite family
of sequences.

Assume that there exists a family of sequences
$S_1, S_2, \ldots$ such that
for any positive $c$ there exists an $i$ such that
$\LDC_W(S_i) \geq \ALG_W(S_i) + c$.
Then we prove that there exists another family of sequences
$S_1', S_2', \ldots$ such that
for any positive $c'$ there exists an $i$ such that
$\ALG_W(S_i') \geq \LDC_W(S_i') + c'$.

This establishes that if $\ALG{}$ performs more than a constant
better on its worst permutations of some family of sequences than \LDC{}
does on its worst permutations, then there exists a family
where $\LDC{}$ has a similar advantage over \ALG{}, which implies
that the algorithms are not comparable.

Now assume that we are given a constant $c$.
Since we must find a value greater than any constant to establish
the result, we may assume without loss of generality that $c$ is large enough
that $3dc \geq \frac{3d+1}{d-1}(3d)+3d$.

Consider a sequence $S$ from the family $S_1, S_2, \ldots$ such that
$\LDC_W(S) \geq \ALG_W(S) + 3dc$.
{}From $S$ we create a member $S'$ of the family $S_1', S_2', \ldots$
such that $\ALG_W(S') \geq \LDC_W(S') + c$.


The idea behind the construction is to have the cruel adversary
against \ALG{} choose requests
from the multiset defined from $S$ as in the definition of canonical
worst orderings. This process continues until
the cruel adversary has used all of either the $A$s, $B$s, or $C$s in the
multiset, resulting in a sequence $S'$.
If the remaining requests from the multiset are
concatenated to $S'$ in any order, this creates a permutation of $S$.
The performance of \ALG{} on this permutation must be at least as
good as its performance on its worst ordering of $S$.

We now consider the performance of \LDC{} and \ALG{} on $S'$
and show that \LDC{} is strictly better.

Let $n_A'$, $n_B'$, and $n_C'$ denote the number of
$A$s, $B$s, and $C$s in $S'$, respectively.

Let
$p=\MIN{\FLOOR{\frac{n_A'}{d}},
        \FLOOR{\frac{n_B'}{d+1}},
        n_C'}$.

By Lemma~\ref{a-LDC-worst}, the cost of \LDC{} on its canonical worst 
ordering of $S'$ is at most $p(4d)+3d$.

The cost of \ALG{} is $2dn_C'+n_A'+n_B'-n_C'$, since
every time there is a request for $C$, this is because a server
in the step before moved away from $C$. These two moves combined
have a cost of $2d$. Every request to an $A$ or a $B$ has cost one,
except for the request to $B$ immediately followed by a
request to $C$, which has already been counted in the $2dn_C'$ term.
A similar argument shows that \LDC{}'s cost is bounded from above
by the same term.

Assume first that $\frac{n_A'}{d}=\frac{n_B'}{d+1}=n_C'$.
Then $S'$ can be permuted so that it is a prefix
of \LDC{}'s canonical worst ordering
on $S$ (see Lemma~\ref{a-LDC-worst} with $a=1$).
Since, by construction, we have run out of either $A$s, $B$s, or $C$s
(that is, one type is missing from $S$ minus $S'$ as multisets),
\LDC{}'s cost on its worst ordering of $S$ is at most its cost on its
worst ordering on $S'$ plus $3d$.
Thus, $\LDC_W(S) \geq \ALG_W(S) + c$ does not hold in this case,
so we may assume that these values are not all equal.

We compare \LDC{}'s canonical worst orderings of $S$ and $S'$. For both
sequences,
the form is as in Lemma~\ref{a-LDC-worst}, with $a=1$. Thus, for $S'$ the
form is $((BA)^d BC)^p X$, and for $S$, it is
$((BA)^d BC)^{p+l} Y$ for some nonnegative integer
$l$. The sequence $X$ must contain
all of the $A$s, all of the $B$s or all of the $C$s contained
in $((BA)^d BC)^l$, since after this the cruel adversary
has run out of something. Thus, it must contain at least $ld$ $A$s,
$l(d+1)$ $B$s or $l$ $C$s. The extra cost that \LDC{} has
over \ALG{} on $S$  is at most its cost on $((BA)^d BC)^l Y$
minus cost $ld$ for the $A$s, $B$s or $C$s contained in $X$, so
at most
$l(2d+2d)+3d-ld=3dl+3d$.
Thus, $\LDC_W(S) - \ALG_W(S) \leq 3dl+3d$.

Since we could assume that not all of $\frac{n_A'}{d}$, $\frac{n_B'}{d+1}$,
and $n_C'$ were equal, we have the following cases:

Case $n_A'>dp$:
\LDC{}'s cost on $S'$ is at most the cost of \ALG{} minus $(n_A'-dp)$
plus $3d$.

Case $n_B'>(d+1)p$:
\LDC{}'s cost on $S'$ is at most the cost of \ALG{} minus $(n_B'-(d+1)p)$
plus $3d$.

Case $n_C'>p$:
\LDC{}'s cost on $S'$ is at most the cost of \ALG{} minus $(2d-1)(n_C'-p)$
plus 1.

Thus, $\ALG_W(S') - \LDC_W(S') \geq dl-3d$.

{}From the choice of $c$,
the definition of the $S_i$ family, and the bound on the
difference between the two algorithms on $S$, we find that
\[\frac{3d+1}{d-1}(3d)+3d\leq3dc\leq\LDC_W(S) - \ALG_W(S)\leq
3dl+3d\]

Thus, $l\geq \frac{3d+1}{d-1}$, which implies the following:
%
\[
l \geq \frac{3d+1}{d-1}
\;\Leftrightarrow\;
ld-3d \geq l+1
\;\Leftrightarrow\;
ld-3d \geq \frac{3dl+3d}{3d}
\]

Now,
\[
\ALG_W(S') - \LDC_W(S') \geq l{d}-3d
\geq \frac{3dl+3d}{3d}
\geq \frac{3dc}{3d}
= c.
\]

Finally, to show that \LDC{} and \ALG{} are not weakly comparable
in \ALG{}'s favor, we show that $c_{\text{u}}(\LDC,\ALG)$ is bounded.
Since the Competitive Ratio of \LDC{} is $2$,
for any algorithm \ALG{} and  any sequence $I$, there is a constant $b$
such that
$\LDC(I_{\LDC})\leq 2\OPT(I_{\LDC}) +b
              \leq 2\ALG(I_{\LDC}) +b 
              \leq 2\ALG(I_{\ALG}) +b$.
Thus, $c_{\text{u}}(\LDC,\ALG)\leq 2$.
\end{proof}


Considering the request sequence as constructed by the cruel
adversary against some algorithm \ALG{}, it consists
of a first part, where the cruel adversary keeps requesting
unoccupied points, and a second part which are all remaining
requests.
The proof of optimality depends on \LDC{} performing as well
as any algorithm on the first part,
and having constant cost on the second part.
Since the first part consists of subsequences where \ALG{}
at some point has a server pulled away from $C$ and then right
back again, it is easy to see that if the distribution of $A$s,
$B$s, and $C$s in those subsequences is different from the
distribution in a canonical worst ordering for \LDC{}, \LDC{} will
simply do better. On the second part, if there are only requests to
two points, \LDC{} will have its two servers on those two points
permanently after a cost of at most $3d$.
Thus, similar proofs will show that
\aLDC{} and \BAL{} are also optimal algorithms,
whereas \Greedy{} is not.

In the definitions of \LDC{} and \BAL{} given in
Section~\ref{preliminaries},
different decisions are made as to which server to use in cases of
ties. In \LDC{} the server which is really closer is moved in
the case of a tie (with regard to virtual distances from the point requested).
The rationale behind this is that the server which
would have the least cost is moved.  In \BAL{} the server which
is further away is moved to the point. The rationale behind this is
that, since $d>1$, when there is a tie, the total cost for the closer 
server is already significantly higher than the total cost for the
other, so moving the server which is further away evens out how
much total cost they have. With these tie-breaking
decisions, the two algorithms behave very similarly.

\begin{theorem}
\label{th:LDC_BAL}
\LDC{} and \BAL{} are equivalent for the baby server problem
according to Relative Worst Order Analysis.
\end{theorem}
\begin{proof}
Consider any request sequence $I$.
\LDC{}'s canonical worst ordering has a prefix of the form
$((BA)^dBC)^k$, while \BAL{}'s canonical worst
ordering has a prefix of the form
\[(BA)^{\FLOOR{\frac{d}{2}}}BC((BA)^dBC)^{k'},\]
such that the remaining parts have constant costs.
These prefixes of \LDC{}'s and \BAL{}'s canonical worst orderings of $I$ 
are identical, except for the constant cost sequence
that \BAL{} starts with.
This also leads to a small constant cost difference at the end.
Thus, their performance
on their respective worst orderings will be identical up to an additive
constant.
\end{proof}

\section{Concluding Remarks}
The purpose of quality measures is to give information
for use in practice, to choose the best algorithm for
a particular application. What properties should such quality measures
have?

First, it may be desirable that  if one algorithm
does at least as well as another on every sequence, then
the measure decides in favor of the better algorithm. This
is especially desirable if the better algorithm does significantly
better on important sequences.
Bijective Analysis and Relative Worst
Order Analysis have this property, but
Competitive Analysis, the Max/Max Ratio, and the Random Order
Ratio do not.
This was seen here in
the lazy vs.\ non-lazy version of Double Coverage for the baby 
server problem
(and the more general metric $k$-server problem).
Similar results have been presented previously for the paging
problem---LRU vs.\ FWF and look-ahead vs.\ no look-ahead.
See~\cite{BFL07} for these results under Relative Worst Order Analysis
and~\cite{ADLO07} for Bijective Analysis. It appears that analysis
techniques that avoid a comparison to \OPT{} have an advantage in
this respect.

Secondly, it may be desirable that, if one algorithm does
unboundedly worse than another on some important families of sequences, the
quality measure reflects this. For the baby server problem, \Greedy{}
is unboundedly worse than \LDC{} on all families of sequences which
consist mainly of alternating requests to the closest two points.
This is reflected in Competitive
Analysis, the Random Order Ratio, and Relative
Worst Order Analysis, but not by the Max/Max Ratio or
Bijective Analysis.
Similarly, according to Bijective 
Analysis, LIFO and LRU are equivalent for paging, but 
LRU is often significantly better
than LIFO, which keeps the first $k-1$ pages it sees in cache forever.
In both of these cases, Relative Worst Order Analysis
says that the algorithms are weakly comparable in favor of
the ``better'' algorithm.

Another desirable property would be ease of computation for many
different problems, as with Competitive Analysis and Relative Worst Order
Analysis.  It is not clear that the Random Order Ratio or
Bijective Analysis have this property.

In this paper, we have initiated a systematic comparison of
quality measures for online algorithms. We hope this will inspire
researchers to similarly investigate a range of online problems
to enable the community to draw stronger conclusions on the relative
strengths of the different measures.

\section*{Acknowledgements}
The authors would like to thank Christian Kudahl for calling
their attention to two oversights in a previous version of this
paper, one in the definition of the lazy version of an algorithm, and another
in the modified definition of the Random Order Ratio.

\bibliography{ref}
\bibliographystyle{plain}

\end{document}